\documentclass[pre,aps,twocolumn]{revtex4} 
\usepackage{psfig}
\usepackage{amsmath}
\usepackage{graphicx}

\newcommand{\ti}{\textit}
\newcommand{\be}{\begin{equation}}
\newcommand{\ee}{\end{equation}}
\newcommand{\bea}{\begin{eqnarray}}
\newcommand{\eea}{\end{eqnarray}}

\begin{document}

\title{Artificial sequences and complexity measures}

\author{Andrea Baronchelli$^1$\footnote{baronka@pil.phys.uniroma1.it}, 
Emanuele Caglioti$^2$\footnote{caglioti@mat.uniroma1.it} and
Vittorio Loreto$^1$\footnote{loreto@roma1.infn.it}}

\affiliation{$^1$''La Sapienza'' University, Physics Department, P.le
A. Moro 5, 00185 Rome, Italy and INFM-SMC, Unit\`a di Roma1.}

\affiliation{$^2$''La Sapienza'' University, Mathematics Department, P.le
A. Moro 5, 00185 Rome, Italy}

\begin{abstract}

In this paper we exploit concepts of information theory to address the
fundamental problem of identifying and defining the most suitable
tools to extract, in a automatic and agnostic way, information from a
generic string of characters. We introduce in particular a class of
methods which use in a crucial way data compression techniques in
order to define a measure of remoteness and distance between pairs of
sequences of characters (e.g. texts) based on their relative
information content. We also discuss in detail how specific features
of data compression techniques could be used to introduce the notion
of {\em dictionary} of a given sequence and of {\em Artificial Text}
and we show how these new tools can be used for information extraction
purposes. We point out the versatility and generality of our method
that applies to any kind of corpora of character strings independently
of the type of coding behind them. We consider as a case study
linguistic motivated problems and we present results for automatic
language recognition, authorship attribution and self
consistent-classification.

\end{abstract}

\maketitle

\section{Introduction}

One of the most challenging issues of recent years is presented by the
overwhelming mass of available data. While this abundance of
information and the extreme accessibility to it represents an
important cultural advance, it raises on the other hand the
problem of retrieving relevant information. Imagine entering the
largest library in the world, seeking all relevant documents on your
favorite topic. Without the help of an efficient librarian this would
be a difficult, perhaps hopeless, task. The desired references would
likely remain buried under tons of irrelevancies. Clearly the need
for effective tools for information retrieval and analysis is becoming
more urgent as the databases continue to grow.

First of all let us consider some among the possible sources of
information. In nature many systems and phenomena are often
represented in terms of sequences or strings of characters. In
experimental investigations of physical processes, for instance, one
typically has access to the system only through a measuring device
which produces a time record of a certain observable, i.e. a sequence
of data. On the other hand other systems are intrinsically described
by string of characters, e.g. DNA and protein sequences, language.

When analyzing a string of characters the main question is to extract
the information it brings. For a DNA sequence this would correspond,
for instance, to the identification of the subsequences codifying the
genes and their specific functions. On the other hand for a written
text one is interested in questions like recognizing the language in
which the text is written, its author or the subject treated.

One of the main approach to this problems, the one we address in this
paper, is that of information theory (IT)~\cite{shannon,zurek} and in
particular the theory of data compression.

In a recent letter~\cite{bcl} a method for context recognition and
context classification of strings of characters or other equivalent
coded information has been proposed. The remoteness between two
sequences $A$ and $B$ was estimated by zipping a sequence $A+B$
obtained by appending the sequence $B$ after the sequence $A$ and
exploiting the features of data compression schemes like {\em gzip}
(whose core is provided by the Lempel-Ziv 77 (LZ77)
algorithm~\cite{LZ77}). This idea was used for authorship attribution
and, by defining a suitable distance between sequences, for languages
phylogenesis.

The idea of appending two files and zipping the resulting file in order to
measure the remoteness between them had been previously proposed by
Loewenstern et al.~\cite{loewenstern} (using {\em zdiff} routines) who
applied it 
to the analysis of DNA sequences, and by Khmelev~\cite{khmelev} who
applied the method to authorship attribution. Similar methods have been
proposed by Juola~\cite{juola}, Teahan~\cite{teahan} and
Thaper~\cite{thaper}.

In this paper we extend the analysis of~\cite{bcl} and we describe in
details the methods to define and measure the remoteness (or
similarity) between pairs of sequences based on their relative
informatic content. We devise in particular, without loss of
generality with respect to the nature of the strings of characters, a
method to measure this {\em distance} based on data-compression
techniques.

The principal tool for the application of these methods is the LZ77
algorithm, which, roughly speaking, achieves the compression of a file
exploiting the presence of repeated subsequences. We introduce (see
also~\cite{phys_a}) the notion of \ti{dictionary} of a sequence,
defined as the set of all the repeated substrings found by LZ77 in a
sequential parsing of a file, and we refer to these substrings as
dictionary's \ti{words}. Besides being of great intrinsic interest,
every dictionary allows for the creation of \ti{Artificial texts} (AT)
obtained by the concatenation of random extracted words. In this paper
we discuss how comparing AT, instead of the original sequences, could
represent a valuable and coherent tool for information extraction to
be used in very different domains. We then propose a general AT
comparison scheme (ATC) and show that it yields to remarkable results
in experiments.
 
We have chosen for our tests some textual corpora and we have
evaluated our method on the basis of the results obtained on some
linguistic motivated problems. Is it possible to automatically
recognize the language in which a given text is written? Is it
possible to automatically guess the author and the subject of a given
text? And finally is it possible to define methods for the automatic
classification of the texts of a given corpus?

The choice of the linguistic framework is justified by the fact that
this is a field where anybody could be able to judge, at least
partially, about the validity and the relevance of the results. Since
we are introducing techniques for which a benchmark does not exist it
is important to check their validity with known and controlled
examples.  This does not mean that the range of applicability is
reduced to linguistics. On the contrary the ambition is to provide
physicists with tools which could parallel other standard tools to
analyze strings of characters.

In this perspective it is worthwhile recalling here some of the last
developments of sequence analysis in physics related problems. A first
field of activity~\cite{paper-yak,fukuda} is that of segmentation
problems, i.e. cases in which a unique string must be partitioned into
subsequences according to some criteria to identify discontinuities in
its statistical properties. A classical example is that of the
separation of coding and non-coding portions in the DNA but the
analysis of genetic sequences in general represents a very rich source
of segmentation problems (see, for instance,
\cite{mantegna,grosse,azad,phys_a}).

\noindent A more recent area is represented by the use of data compression
techniques to test specific properties of symbolic
sequences. In~\cite{kennel}, the technology behind adaptive dictionary
data compression algorithms is used in a suitable way (which is very
close to our approach) as an estimate of reversibility of time series,
as well as a statistical likelihood test. Another interesting field is
related to the problem of the generation of random
numbers. In~\cite{mertens} it is outlined the importance of suitable
measures of conditional entropies in order to check the real level of
randomness of random numbers, and an entropic approach is used to
discuss some random number generator shortcomings (see
also~\cite{vulpio}).

\noindent Finally, another area of interest is represented by the use of data
compression techniques to estimate entropic quantities (e.g. Shannon
entropy, Algorithmic Complexity, Kullback-Leibler divergence etc.)
Even though not new this area is still
topical~\cite{grassberger2,shurmann}. A specific application that has
generated an interesting debate has been drawn about the analysis of
electroencephalograms of epilepsy patients~\cite{enc1,enc2,enc3}.  In
particular in these paper it is argued that measures like the
Kullback-Leibler divergence could be used to spot information in
medical data. The debate is wide open.

The outline of the paper is as follows. In section II, after a short
theoretical introduction, we recall how data compression techniques
could be used to evaluate entropic quantities. In particular we recall
the definition of the LZ77~\cite{LZ77} compression algorithm and we
address the problem of using it to evaluate quantities like the
relative entropy between two generic sequences as well as to define a
suitable distance between them. In section III we introduce the
concept of Artificial Text (AT) and present a method for information
extraction based on Artificial Text comparison. Sections IV and V are
devoted to the results obtained with our method in two different
contexts: the recognition and extraction of linguistic features
(sec. IV) and the self-consistent classification of large corpora
(sec. V).  Finally section VI is devoted to the conclusions and to a
short discussion about possible perspectives.

\section{Complexity Measures and Data Compression}

Before entering in the details of our method let us briefly recall the
definition of entropy of a string. Shannon's definition of information
entropy is indeed a probabilistic concept referring to the source 
emitting strings of characters.

Consider a symbolic sequence $(\sigma_1\, \sigma_2\, \dots)$, where
$\sigma_t$ is the symbol emitted at time $t$ and each $\sigma_t$ can
assume one of $m$ different values. Assuming that the sequence is
stationary we introduce the $N-$block entropy:
\begin{equation}
H_N=-\sum_{\{W_N \}} p(W_N) \ln p(W_N)
\end{equation}
\noindent where $p(W_N)$ is the probability of the $N$-word
$W_N=(\sigma_t\,\sigma_{t+1}\,\dots \, \sigma_{t+N-1})$, and $\ln = \log_e$.
The differential entropies

\begin{equation}
h_N=H_{N+1}-H_N
\end{equation}

\noindent have a rather obvious meaning: $h_N$ is the average information
supplied by the $(N+1)$-th symbol, provided the $N$ previous ones are
known. Noting that the knowledge of a longer past history cannot
increase the uncertainty on the next outcome, one has that $h_N$
cannot increase with $N$ i.e. $h_{N+1} \leq h_N$. 
With these definitions the Shannon entropy for an ergodic stationary process
is defined as:

\begin{equation} 
\label{h_definition}
h=\underset{N \to \infty}{\lim} h_N = \underset{N \to \infty}{\lim}
\frac{H_N}{N}.
\end{equation}

It is easy to see that for a $k$-th order Markov process (i.e. such
that the conditional probability to have a given symbol only depends
on the last $k$ symbols,
$p(\sigma_t|\sigma_{t-1}\,\sigma_{t-2},\dots) =
p(\sigma_t|\sigma_{t-1}\,\sigma_{t-2},\dots,\sigma_{t-k})$, 
then  $h_N=h$ for $N \geq k$.

The Shannon entropy $h$ measures the average amount of information per
symbol and it is an estimate of the ``surprise'' the source emitting
the sequence reserves to us. It is remarkable the fact that, under
rather natural assumptions, the entropy $H_N$ apart from a
multiplicative factor, is the unique quantity which characterizes the
``surprise'' of the $N$-words~\cite{K57}.  Let's try to explain in
which sense entropy can be considered as a measure of a surprise.
Suppose that the surprise one feels upon learning that an event E has
occurred depends only on the probability of E. If the event occurs
with probability 1 (sure) our surprise in its occurring will be
zero. On the other hand if the probability of occurrence of the event
E is quite small our surprise will be proportionally large. For a
single event occurring with probability $p$ the surprise is
proportional to $\ln p$.  Let's consider now a random variable $X$,
which can take $N$ possible values $x_1,...,x_N$ with probabilities
$p_1,...,p_N$, the expected amount of surprise we shall receive upon
learning the value of $X$ is given precisely by the entropy of the
source emitting the random variable $X$, i.e.  $- \sum p_i \ln p_i$.

The definition of entropy is closely related to a very old problem,
that of transmitting a message without loosing information, i.e. the
problem of the efficient encoding~\cite{welsh}.

A good example is the Morse code. In the Morse code a text is encoded
with two characters: line and dot. What is the best way to encode the
characters of the English language (provided one can define a source
for English) with sequences of dots and lines?  The idea of Morse was
to encode the more frequents characters with the minimum numbers of
characters. Therefore the $e$ which is the most frequent English
letter is encoded with one dot ($\cdot$), while the letter $q$ is
encoded with three lines and one dot ($- - \cdot -$).

The problem of the optimal coding for a text (or an image or any other
kind of information) has been enormously studied.  In particular
Shannon~\cite{shannon} showed that there is a limit to the possibility
to encode a given sequence. This limit is the entropy of the sequence.

This result is particularly important when the aim is the measure of
the information content of a single finite sequence, without any
reference to the source that emitted it. In this case the reference
framework is the Algorithmic Complexity Theory and the basic concept is
Chaitin - Kolmogorov entropy or Algorithmic Complexity
(AC)~\cite{k65,Ch66,Ch90,S64}: {\em the entropy of a string of
characters is the length (in bits) of the smallest program which
produces as output the string and stops afterwords}.  This definition
is really abstract. In particular it is impossible, even in principle,
to find such a program and as a consequence the algorithmic complexity
is a non computable quantity.  This impossibility is related to the
halting problem and to Godel's theorem~\cite{livit}.

It is important to recall how it exists a rather important relation
between the Algorithmic Complexity $K_N(W_N)$ of a sequence $W_N$ of
$N$ characters and $H_N$:

\begin{equation}
\frac{1}{N}\langle K_N \rangle=
\frac{1}{N}\sum_{W_N} K_N(W_N)P(W_N) 
\xrightarrow[N \to \infty]{} \frac{h}{\ln 2}
\end{equation}
where $K_N$ is the binary length of the shorter program needed to
specify the sequence $W_N$.

As a consequence it exists a relation between the maximum compression
rate of a sequence $(\sigma_1\,\sigma_2\,\dots)$ expressed in an
alphabet with $m$ symbols, and $h$. If the length $N$ of the sequence
is large enough, then it is not possible to compress it into another
sequence (with an alphabet with $m$ symbols) whose size is smaller
than $N h/\ln m$. Therefore, noting that the number of bits needed for
a symbol in an alphabet with $m$ symbol is $\ln m$, one has that the
maximum allowed compression rate is $h/\ln m$~\cite{shannon}.

Though the maximal theoretical limit of the Algorithmic Complexity is
not achievable, there are nevertheless algorithms explicitly conceived
to approach it. These are the file compressors or zippers. A zipper
takes a file and tries to transform it in the shortest possible
file. Obviously this is not the best way to encode the file but it
represents a good approximation of it.

\begin{figure}
\centerline{\psfig{figure=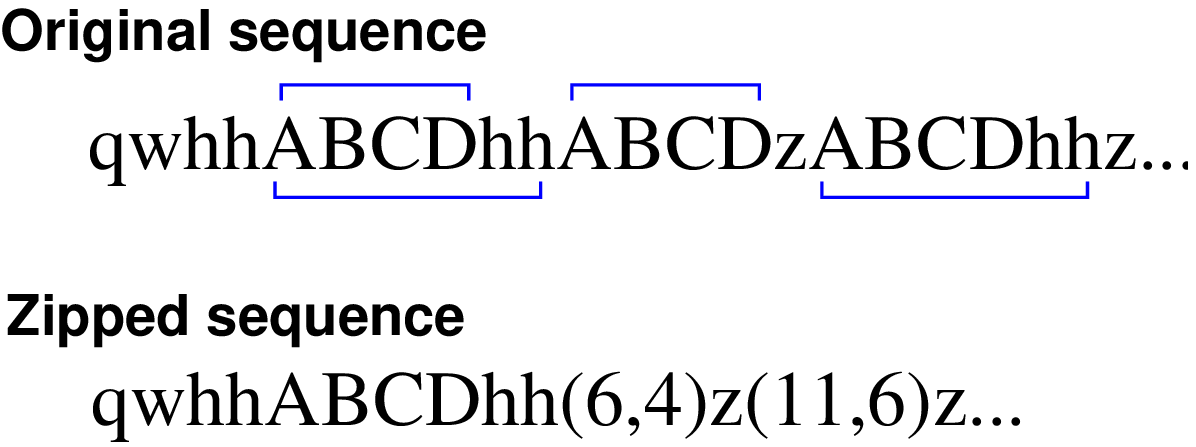,width=8cm,angle=0}}
\vspace{0.3cm}
\caption{{\bf Scheme of the LZ77 algorithm}: The LZ77 algorithm works
sequentially and at a generic step looks in the look-ahead buffer for
substrings already encountered in the buffer already scanned. These
substrings are substituted by a pointer (\ti{d,n}) where \ti{d} is the
distance of the previous occurrence of the same substring and \ti{n}
is its length.Only strings longer than two characters are substituted
in the example.}
\label{lz77fig}
\end{figure}

A great improvement in the field of data compression has been
represented by the Lempel and Ziv algorithm (LZ77)~\cite{LZ77} (used
for instance by $gzip$ and $zip$). It is interesting to briefly recall
how it works (see fig.~\ref{lz77fig}). Let $x=x_1,....,x_N,$ the
sequence to be zipped, where $x_i$ represents a generic character of
sequence's alphabet. The LZ77 algorithm finds duplicated strings in
the input data. The second occurrence of a string is replaced by a
pointer to the previous string given by two numbers: a distance,
representing how far back into the window the sequence starts, and a
length, representing the number of characters for which the sequence
is identical. More specifically the algorithm proceeds sequentially
along the sequence. Let us suppose that the first $n$ characters have
been codified. Then the zipper looks for the largest integer $m$ such
that the string $x_{n+1},...,x_{n+m}$ already appeared in
$x_1,...,x_n$. Then it codifies the string found with a two-number
code composed by: the distance between the two strings and the length
$m$ of the string found. If the zipper does not find any match then it
codifies the first character to be zipped, $x_{n+1}$, with its
name. This eventuality happens for instance when codifying the first
characters of the sequence, but this event becomes very infrequent as
the zipping procedure goes on.

This zipper is asymptotically optimal: i.e. if
it encodes a text of length $L$ emitted by an ergodic source whose
entropy per character is $h$, then the length of the zipped file
divided by the length of the original file tends to $h$ when the
length of the text tends to $\infty$. The convergence to this limit is
slow and the corrections has been shown to behave as $O\left({\log\log
L\over \log L}\right)$~\cite{lz77convergence}.

Usually, in commercial implementations of LZ77 (like for instance
$gzip$), substitutions are made only if the two identical sequences
are not separated by more than a certain number $n_w$ of characters,
and the zipper is said to have a $n_w$-long sliding window. The
typical value of $n_w$ is 32768. The main reason for this restriction
is that the search in very large buffers could be not efficient from
the computational time point of view.

Just to give an example, if one compresses an English text the length
of the zipped file is typically of the order of one fourth of the
length of the initial file. An English file is encoded with $1$ byte
($8$ bits) per character. This means that after the compression the
file is encoded with about $2$ bits per character. Obviously this is
not yet optimal. Shannon with an ingenious experiment showed that the
entropy of the English text is between $0.6$ and $1.3$ bits per
character~\cite{pierce} (for a recent study see~\cite{grassberger2}).

It is well known that compression algorithms represent a powerful tool
for the estimation of the AC or more sophisticated measures of
complexity~\cite{wyner,ziv-merhav,farach,milo,cai} and several
applications have been drawn in several fields~\cite{verdu} from
dynamical systems theory (the connections between Information Theory
and Dynamical Systems theory are very strong and go back all the way
to Kolmogorov and Sinai works~\cite{sinai59,eck85}. For a recent
overview see~\cite{lind,benci_1,boffetta}) to linguistics (an
incomplete list would include
~\cite{bell,nevill-manning,el-yaniv,juola,kontoyiannis,khmelev,teahan,thaper,bcl,sciam}),
genetics
(see~\cite{grumbach,loewenstern,li,li_similarity,menconi,phys_a} and
references therein) and music classification~\cite{cilibrasi_music,londei}.

\subsection{Remoteness between two texts}

It is interesting to recall the notion of relative entropy (or
Kullback-Leibler divergence~\cite{KL,Kullback,cover-thomas}) which is
a measure of the statistical remoteness between two distributions and
whose essence can be easily grasped with the following example. 

Let us consider two stationary zero-memory sources $\cal A$ and $\cal
B$ emitting sequences of $0$ and $1$: $\cal A$ emits a $0$ with probability $p$
and $1$ with probability $1-p$ while $\cal B$ emits $0$ with
probability $q$ and $1$ with probability $1-q$. As already described,
a compression algorithm like LZ77 applied to a sequence emitted by
$\cal A$ will be asymptotically (i.e. in the limit of an available
infinite sequence) able to encode the sequence almost optimally,
i.e. coding on average every character with $-p \, \log_2 p -(1-p) \,
\log_2(1-p)$ bits (the Shannon entropy of the source). This optimal
coding will not be the optimal one for 
the sequence emitted by $\cal B$. In particular the entropy per
character of the sequence emitted by $\cal B$ in the coding optimal
for $\cal A$ (i.e. the cross-entropy per character) will be $-q
\,log_2 p - (1-q) \, log_2 (1-p)$ while the entropy per character of
the sequence emitted by $\cal B$ in its optimal coding is $-q \, log_2
q - (1-q) \, log_2 (1-q)$. The number of bits per character waisted to
encode the sequence emitted by $\cal B$ with the coding optimal for
$\cal A$ is the relative entropy of $\cal A$ and $\cal B$,

\begin{equation}
d({\cal A}\vert \vert {\cal B})= -q \, log_2 \frac{p}{q} - (1-q) \, log_2
\frac{1-p}{1-q}
\end{equation}

A linguistic example will help to clarify the situation: transmitting
an Italian text with a Morse code optimized for English will result in
the need of transmitting an extra number of bits with respect to
another coding optimized for Italian: the difference is a measure of
the relative entropy between, in this case, Italian and English
(supposing the two texts are each one archetypal representations of
their Language, which is not).

We should remark that the relative entropy is not a distance (metric)
in the mathematical sense: it is neither symmetric, nor does it
satisfy the triangle inequality. As we shall see below, in many
applications, such as phylogenesis, it is crucial to define a true
metric that measures the actual distance between sequences.

There exist several ways to measure the relative entropy (see for
instance~\cite{wyner,ziv-merhav,cai}). One possibility is of course
to follow the recipe described in the previous example: using the
optimal coding for a given source to encode the messages of another
source. 

Here we follow the approach recently proposed in~\cite{bcl} which is
similar to the approach by Ziv and Merhav~\cite{ziv-merhav}.  In
particular in order to define the relative entropy between two sources
$\cal A$ and $\cal B$ we consider a sequence $A$ from the source $\cal
A$ and a sequence $B$ from the source $\cal B$. We now perform the
following procedure. We create a new sequence $A+B$ by appending $B$
after $A$ and use the LZ77 algorithm or, as we shall see below, a
modified version of it.

In~\cite{paper-yak} it has been studied in detail what happens when a
compression algorithm tries to optimize its features at the interface
between two different sequences $A$ and $B$ while zipping the sequence
$A+B$ obtained by simply appending $B$ after $A$. It has been shown in
particular the existence of a scaling function ruling the way the
compression algorithm learns a sequence $B$ after having compressed a
sequence $A$. In particular it turns out that it exists a crossover
length for the sequence $B$, given by

\begin{equation}
L_B^* \simeq L_A^\alpha
\label{learning}
\end{equation}

\noindent with $\alpha=\frac {h(B)}{h(B)+ d(B \vert \vert A)}$. This
is the length below which the compression algorithm does not learn
the sequence $B$ (measuring in this way the cross entropy between $A$
and $B$) and above which it learns $B$, i.e. optimizes the compression
using the specific features of $B$.

This means that if $B$ is short enough (shorter than the crossover
length), one can measure the relative entropy by zipping the sequence
$A+B$ (using $gzip$ or an equivalent sequential compression program);
the measure of the length of $B$ in the coding optimized for $A$ will
be $\Delta_{AB}= L_{A+B} - L_{A}$, where $L_{X}$ indicates the length
in bits of the zipped file $X$. The cross entropy per character
between $\cal A$ and $\cal B$ will be estimated by
\begin{equation}
C{(\cal{A}|\cal{B})} = \Delta_{AB} /|B|,
\label{cross-ent}
\end{equation}
where $|B|$ is the length in bits of the uncompressed file $B$.  The
relative entropy $d({\cal A}\vert \vert {\cal B})$ per character
between $\cal A$ and $\cal B$ will be estimated by
\begin{equation}
d{(\cal{A}\vert \vert {\cal B})} = (\Delta_{AB} -
\Delta_{B^{\prime}B})/|B|,
\label{rel-ent}
\end{equation}
where $B^{\prime}$ is a second sequence extracted from the source
$\cal B$ with $|B^{\prime}|$ characters and $\Delta_{B^{\prime}B}
/|B|= (L_{B+B^{\prime}} - L_{B})/|B|$ is an estimate of the
entropy of the source $\cal B$.

If, on the other hand, $B$ is longer than the crossover length we must
change our strategy and implement an algorithm which does not zip the
$B$ part but simply ``reads'' it with the (almost) optimal coding of
part $A$. In this case we start reading sequentially file $B$ and
search in the look-ahead buffer of $B$ for the longest sub-sequence
already occurred {\bf only} in the $A$ part. This means that we do not allow
for searching matches inside $B$ itself. As in the usual LZ77, every
matching found is substituted with a pointer indicating where, in $A$,
the matching subsequence appears and its length. This method allows us
to measure (or at least to estimate) the cross-entropy between $B$ and
$A$, i.e. $C{(\cal {A}|{\cal B})}$.

Before proceeding let us briefly discuss which difficulties one could
experiment in the practical implementation of the methods described in
this section. First of all in practical applications the sequences to
be analyzed can be very long and their direct comparison can then be
problematic due to finiteness of the window over which matching can be
found. Moreover in some applications one is interested in estimating
the self-entropy of a source, i.e. $C{(\cal {A}|\cal {A})}$ in
a more coherent framework. The estimation of this quantity is
necessary to calculate the relative-entropy between two sources. In
fact, as we shall see in the next section, even though in practical
applications the simple cross-entropy is often used, there are cases
in which relative entropy is more suitable. The most typical case is
when we need to build a symmetrical distance between two
sequences. One could think to estimate self-entropy comparing, with
the modified LZ77, two portions of a given sequence. This method is
not very reliable since many bias could afflict the results obtained
in this way. For example if we split a book in two parts and try to
measure the cross-entropy between these two parts, the result we would
obtain could be heavily affected by the names of the characters
present in both parts. More importantly, defining the position of the
cut would be completely arbitrary, and this arbitrariness would matter
a lot especially for very short sequences. We shall address this problem
in section III.

\subsection{On the definition of a distance}

In this section we address the problem of defining a distance between
two generic sequences $A$ and $B$. A distance $D$ is an application
that must satisfy three requirements:
\begin{enumerate}
\item positivity: $D_{AB} \geq 0$ ($D_{AB}=0$
iff $A=B$);
\item symmetry: $D_{AB} = D_{BA}$;
\item triangular inequality: $D_{AB} \leq D_{AC} +
D_{CB} \;\; \forall \;C$;
\end{enumerate}

As it is evident the relative entropy $d{(\cal {A}\vert \vert \cal
{B})}$ does not satisfy the last two properties while it is never
negative. Nevertheless one can define a symmetric quantity as follows:
\begin{equation}
P_{AB} = P_{BA} = \frac{C{(\cal{A}|\cal {B})}-C{(\cal{B}|\cal
{B})}}{C{(\cal{B}|\cal {B})}}+\frac{C{(\cal{B}|\cal
{A})}-C{(\cal{A}|\cal {A})}}{C{(\cal{A}|\cal {A})}}
\label{distanza_nostra}
\end{equation} 
We now have a symmetric quantity, but $P_{AB}$ does not satisfies, in
general, the triangular inequality.  In order to obtain a real
mathematical distance we give a prescription according to which this
last property is met. For every pair $A$ and $B$ of sequences, the
prescription writes as:
\begin{eqnarray}
\mbox{ if } \;\; P_{AB} > min_C[P_{AC} + P_{CB}] \;\mbox{ then } \nonumber \\
P_{AB} = min_C[P_{AC} + P_{CB}].
\end{eqnarray}
\noindent By iterating this procedure until for any $A,B,C$ $P_{AB}
\leq P_{AC} + P_{CB},$ we obtain a true distance $D_{AB}$.  In particular the
distance obtained in this way is simply the minimum over all the paths
connecting $A$ and $B$ of the total cost of the path (according to $P_{AB}$):
i.e.
\begin{equation}
D_{AB}=\min_{\{N\geq 2\}}\min_{\{X_1,...,X_N:X_1=A,X_N=B\}} 
\sum_{k=0}^{N-1}P_{X_kX_{k+1}}.
\end{equation}

Also it is easy to see that $D_{AB}$ is the maximal distance not
larger than $P_{A,B}$ for any $A,B$, where we have considered the
partial ordering on the set of distances: $P \geq P^{\prime}$ if
$P_{AB}\geq P^{\prime}_{AB}$, for all pairs $A,B$.

Obviously this is not an a-priori distance. The distance between $A$
and $B$ depends, in principle, on the set of files we are considering.

In all our tests with linguistic texts the triangle condition was
always satisfied without the need to have recourse to the above
mentioned prescription. However there are cases in other contexts,
like, for instance, genetic sequences, in which could be necessary to
force the triangularization procedure described above.

An alternative definition of distance can be given considering
\begin{equation}
R_{AB} = \sqrt{P_{AB}},
\end{equation}
where the square root must be taken before forcing the
triangularization. The idea of using $R_{AB}$ is suggested by the fact
that as $A$ and $B$ are very close sources then $P_{AB}$ is of the
order of the square of their ``difference''.  Let us see this in a
concrete example where the distance between the two sources is very
small.  Suppose having two sources $\cal A$ and $\cal B$ which can
emit sequences of $0$ and $1$.  Let $\cal A$ emit a $0$ with a
probability $p$ and $1$ with the complementary probability $1-p$. Now
let the source $\cal B$ emit a $0$ with a probability $p +\epsilon$
and a $1$ with a probability $1- (p + \epsilon)$, where $\epsilon$ is
an infinitesimal quantity. In this situation it can be easily shown
that the relative entropy between $ \cal A$ and $ \cal B$ is
proportional to $\epsilon^2$ and, of course, $P_{AB}$ is then
proportional to the same quantity. Taking the square root of $P_{AB}$
is then simply requiring that, if two sources have a distribution of
probability that differs for a small $\epsilon $, their distance must
be of the order of $\epsilon $ instead of being reduced to the
$\epsilon^2$ order.

It is important to recall that an earlier and rigorous definition of
an unnormalized distance between two generic strings of characters has
been proposed in~\cite{bennett} in terms of the Kolmogorov Complexity
and of the Conditional Kolmogorov Complexity~\cite{livit} (see below
for the definition).

A normalized version of this distance has been proposed
in~\cite{li_similarity,cilibrasi}. In particular Li et al. define

\be
d_K(x,y)= \frac{max(K(x|y),K(y|x))}{max(K(x),K(y))}
\label{livitdist}
\ee

where the subscript $K$ refers to its definition in terms of the
Kolmogorov complexity. $K(x|y)$ is the conditional Kolmogorov
Complexity defined as the length of the shortest program to compute
$x$ if $y$ is furnished as an auxiliary input to the computation, and
$K(x)$ and $K(y)$ are the Kolmogorov complexities of strings $x$ and
$y$, respectively. The distance $d_K(x,y)$ is symmetrical and it is
shown to satisfy the identity axiom up to a precision $d_K(x,x) =
O(1/K(x))$ and the triangular inequality $d_K(x,y) <= d_K(y,z)+d_K(z,y)$ up
to an additive term $O(1/max(K(x),K(y),K(z)))$.

The problem with this distance is the fact that it is defined in terms
of the Conditional Kolmogorov Complexity which is an uncomputable
quantity and its computation is performed in an approximate way.

In particular what is important is that the specific procedure
(algorithm) used to approximate this quantity, which is indeed a well
defined mathematical operation, defines a true distance.  In the
specific case of the distance $d_K(x,y)$ defined
in~\cite{li_similarity} the authors approximate this distance by the
so-called Normalized Compression Distance

\be 
NCD(x,y) = \frac{C(xy) - min(C(x),C(y))}{max (C(x),C(y))} 
\label{NCD}
\ee 

\noindent where $C(xy)$ is the compressed
size of the concatenation of $x$ and $y$, and $C(x)$ and $C(y)$ denote
the compressed size of $x$ and $y$, respectively.  Then this
quantities are approximated in a suitable way by using real world
compressors.

It is important to remark how it exists a discrepancy between the
definition~\ref{livitdist} and its actual approximate
computation~\ref{NCD}.

We discuss here in some details the case of the LZ77 compressor.
Using the results presented in Sect.IIA, one obtains that, if the
length of $y$ is small enough (see expression~\ref{learning}),
$NCD(x,y)$ is actually estimating the cross-entropy between $x$ and
$y$. The cross-entropy is not a distance since it does not satisfy the
identity axiom, it is not symmetrical nor it satisfies the triangular
inequality.  In the general case of $y$ being not small, again
following the discussion of Sect.IIA (presented in more details
in~\cite{paper-yak}), one can show that $NCD(x,y)$ is given roughly
(for $L_x$ large enough) by:

\be
1+ \frac{L_x^{\alpha}}{L_y} \frac{d(x||y)}{C(y)}, 
\ee

\noindent where $L_x$ and $L_y$ are the lengths of the $x$ and $y$
files (with $L_y >> L_x^{\alpha}$) and $d(x||y)$ is the relative
entropy rate between $x$ and $y$. Again this estimate does not define
a metric.  Moreover, since $\alpha \leq 1$ one can see that
$NCD(x,y)\rightarrow 1$, independently of the choice of $x$ and $y$
when $L_x$ and $L_y$ tends to infinity.

The discrepancy between the definition of a mathematical distance
based on the Conditional Kolmogorov Complexity and its actual
approximate computation in~\cite{li_similarity} has also been pointed
out in~\cite{kaltchenko}.

Finally it is important to notice that recently Otu and
Sayood~\cite{otu-sayood} have proposed an alternative definition of
distance between two string of characters, which is rigorous and
computable. Their approach is based on the LZ complexity~\cite{LZ76}
of a sequence S which can be defined in terms of the number of steps
required by a suitable production process to generate S. In their very
interesting paper they also give a review on this and correlated
problems.  We do not enter here on the details and we refer the reader
to~\cite{otu-sayood}.

\section{Dictionaries and Artificial Texts}

\begin{table}
\begin{center}
\begin{tabular}{|c|c|c|}
\hline
 {\centering {\sf Frequency}} 
&{\centering {\sf Length}} 
&{\centering {\sf Word}} 
\\\hline
 {\centering {\sf 110}} 
&{\centering {\sf 6}} 
&{\centering {\sf .$\smile$The$\smile$}} 
\\\hline
 {\centering {\sf 107}}
&{\centering {\sf 7}}
&{\centering {\sf in$\smile$the$\smile$}}
\\\hline 
 {\centering {\sf 98}}
&{\centering {\sf 4}}
&{\centering {\sf you$\smile$}}
\\\hline
 {\centering {\sf 94}}
&{\centering {\sf 6}}
&{\centering {\sf .$\smile$But$\smile$}}
\\\hline
 {\centering {\sf 92}}
&{\centering {\sf 9}}
&{\centering {\sf from$\smile$the$\smile$}}
\\\hline
{\centering {\sf 92}}
&{\centering {\sf 5}}
&{\centering {\sf $\smile$very$\smile$}}
\\\hline
{\centering {\sf 91}}
&{\centering {\sf 4}}
&{\centering {\sf one$\smile$}}
\\\hline
\end{tabular}
\vspace{0.1cm}
\caption{{\bf Most frequent LZ77-words found in Moby Dick's text}:
Here we present the most represented word in the dictionary of Moby
Dick. The dictionary was extracted using a 32768 sliding window in
LZ77. The $\smile$ represents the space character.}
\label{diz-freq}
\end{center}
\end{table}

\begin{figure}
\begin{tabular}{c}
\psfig{figure=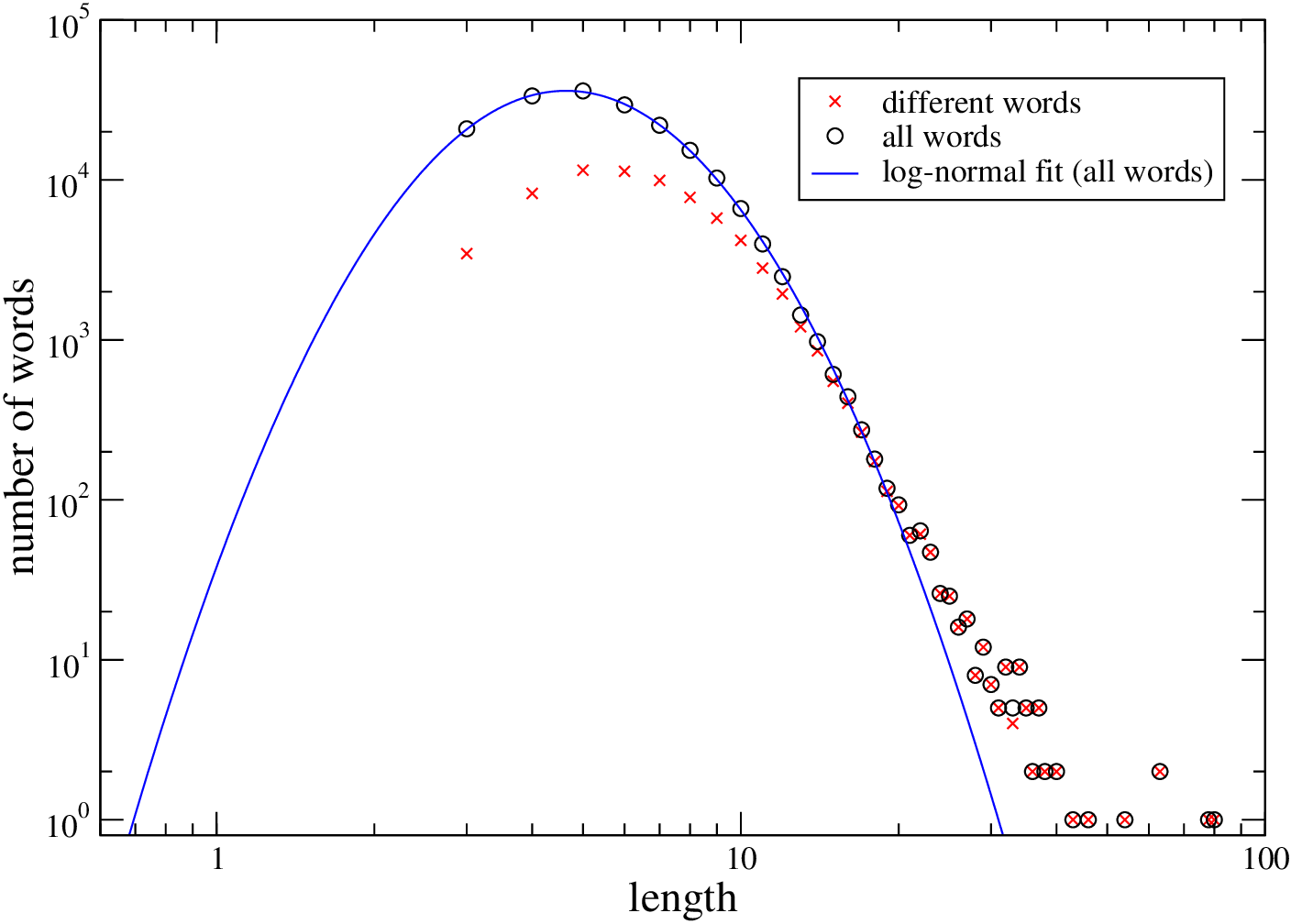,width=8cm,angle=0} \vspace{0.5cm}\\
\psfig{figure=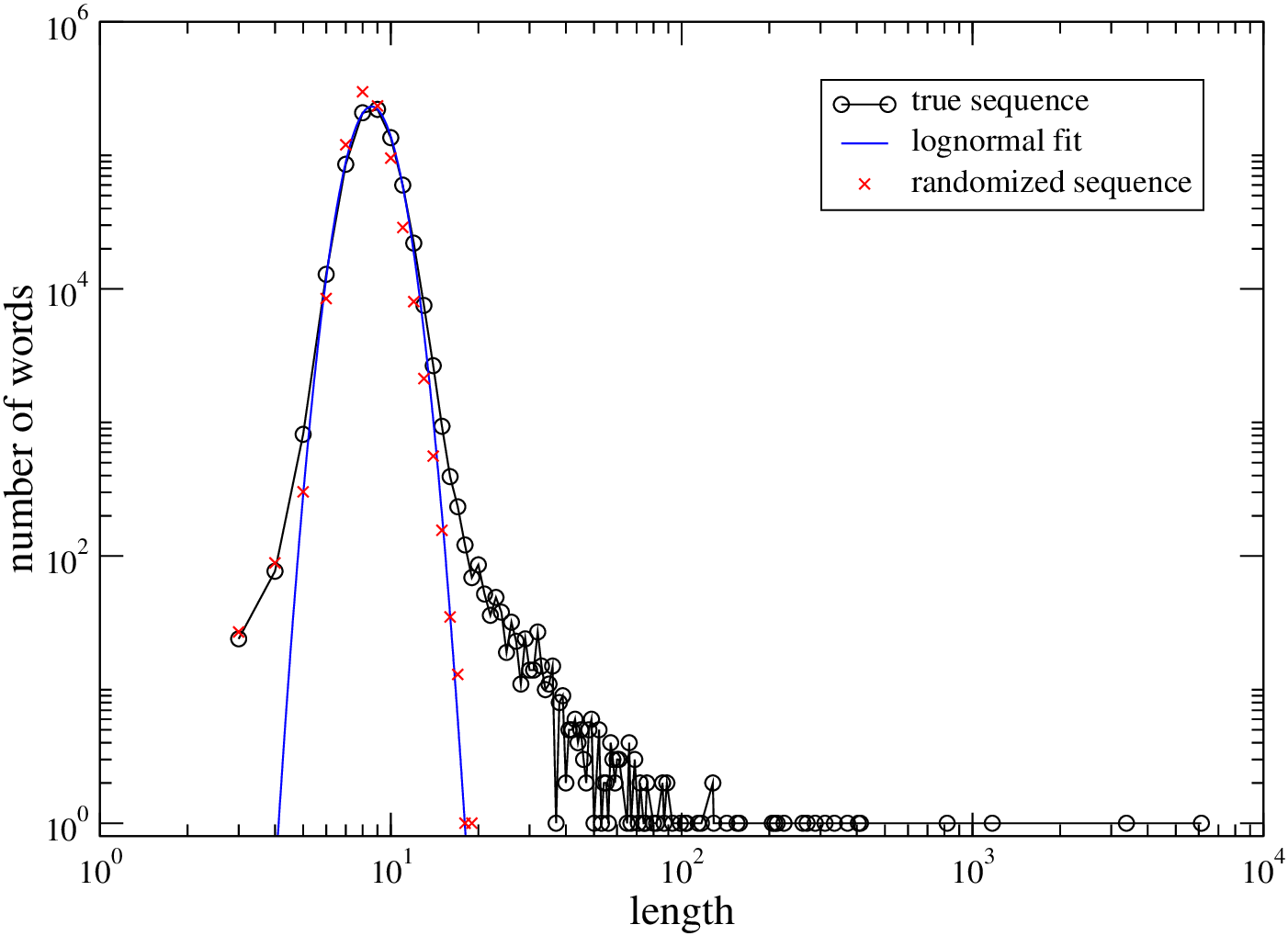,width=8cm,angle=0}\\
\end{tabular}
\caption{{\bf LZ77-Word Distribution} This figure illustrates the
distribution of the LZ77-words found in different strings of
characters. Above: results for the dictionary of {\em Moby Dick} are
shown. In the upper curve several findings of the same word are
considered separately; in the lower curve each different word is
counted only once. It can be shown that the peaks are well fitted by a
log-normal distribution, while there are large deviations from it for
large lengths. Below: words extracted from \textit{Mesorhizobium loti}
bacterium's original and reshuffled DNA sequences are analyzed. The
log-normal curve fits well the whole distribution of words extracted
from the reshuffled string, but is unable to describe the presence of
the long words of the true one.}
\label{fre_len}
\end{figure}

As we have seen LZ77 substitutes sequences of characters with a
pointer to their previous appearance in the text. We now need some
definitions before proceeding. We call \ti{dictionary} of a sequence
the whole set of sub-sequences substituted with a pointer by LZ77, and
we refer to these sequences as dictionary's \ti{words}. As it is
evident from these definitions, a particular word can be present many
times in the dictionary. Finally, we call \ti{root} of a dictionary
the sequence it has been extracted from. It is important to stress how
this dictionary has in principle nothing to do with the ordinary
dictionary of a given language. On the other hand there could be
important similarities between the LZ77-dictionary of a written text
and the dictionary of the Language in which the text is written. As an
example we report in Table~\ref{diz-freq} and Table~\ref{diz-lung} the
most frequent and the longest {\em words} found by LZ77 while zipping
Melville's Moby Dick text. Figure~\ref{fre_len} reports an example of
the frequency-length distribution of the LZ77-words as a function of
their length (for a very similar figure and similar but less complete
dictionary analysis see~\cite{phys_a}).

Beyond their utility for zipping purposes, the dictionaries present an
intrinsic interest since one can consider them as a source for the
principal and more important syntactic structures present in the
sequence/text from which the dictionary originates.

A straightforward application is the possibility to construct
\ti{Artificial Texts}. With this name we mean sequences of characters
build by concatenating words randomly extracted from a specific
dictionary.

\begin{table}
\begin{center}
\begin{tabular}{|c|c|l|}
\hline
{\centering {\sf Frequency}} 
&{\centering {\sf Length}} 
&{\centering {\sf Word}} 
\\\hline
{\centering {\sf 1}} 
&{\centering {\sf 80}} 
&{{\sf ,--$\smile$Such$\smile$a$\smile$funny,$\smile$sporty,$\smile$gamy,}} 
\\
{\centering {\sf }} 
&{\centering {\sf }} 
&{\centering {\sf$\smile$jesty,$\smile$joky,$\smile$hoky-poky$\smile$lad,$\smile$is}}
\\
{\centering {\sf }} 
&{\centering {\sf }} 
&{\centering {\sf $\smile$the$\smile$Ocean,$\smile$oh!$\smile$Th}}
\\\hline
{\centering {\sf 1}} 
&{\centering {\sf 78}} 
&{{\sf ,--$\smile$Such$\smile$a$\smile$funny,$\smile$sporty,$\smile$gamy,}} 
\\
{\centering {\sf }} 
&{\centering {\sf }} 
&{\centering {\sf$\smile$jesty,$\smile$joky,$\smile$hoky-poky$\smile$lad,$\smile$is}}
\\
{\centering {\sf }} 
&{\centering {\sf }} 
&{\centering {\sf $\smile$the$\smile$Ocean,$\smile$oh!$\smile$}}
\\\hline
{\centering {\sf 1}}
&{\centering {\sf 63}}
&{\centering {\sf "$\smile$"I$\smile$look,$\smile$you$\smile$look,$\smile$he$\smile$looks;$\smile$}} 
\\
{\centering {\sf }}
&{\centering {\sf }}
&{\centering {\sf we look,$\smile$ye$\smile$look,$\smile$they look."$\smile$"W}} 
\\\hline
{\centering {\sf 1}}
&{\centering {\sf 63}}
&{\centering {\sf "!$\smile$"I$\smile$look,$\smile$you$\smile$look,$\smile$he$\smile$looks;}} 
\\
{\centering {\sf }}
&{\centering {\sf }}
&{\centering {\sf $\smile$we look,$\smile$ye$\smile$look,$\smile$they look."$\smile$"}} 
\\\hline
{\centering {\sf 1}}
&{\centering {\sf 54}}
&{\centering {\sf repeated$\smile$in$\smile$this$\smile$book,$\smile$that$\smile$the}}
\\
{\centering {\sf }}
&{\centering {\sf }}
&{\centering {\sf the$\smile$skeleton$\smile$of$\smile$the whale}}
\\\hline
{\centering {\sf 1}}
&{\centering {\sf 46}}
&{\centering {\sf .$\smile$THIS$\smile$TABLET$\smile$Is$\smile$erected$\smile$to}}
\\
{\centering {\sf }}
&{\centering {\sf }}
&{\centering {\sf $\smile$his$\smile$Memory$\smile$BY$\smile$HIS$\smile$}}
\\\hline
{\centering {\sf 1}}
&{\centering {\sf 43}}
&{\centering {\sf s$\smile$a$\smile$mild,$\smile$mild$\smile$wind,$\smile$and$\smile$a$\smile$}}
\\
{\centering {\sf }}
&{\centering {\sf }}
&{\centering {\sf mild$\smile$looking$\smile$sky}}
\\\hline
\end{tabular}
\vspace{0.1cm}
\caption{{\bf Longest words in Moby Dick}: Here we present the longest
words in the dictionary of Mody Dick. Each of these words appears only
one time in the dictionary. The dictionary was 
extracted using a 32768 sliding window in LZ77.}
\label{diz-lung}
\end{center}
\end{table}

\vspace{0.2cm}

Each word has a probability of being extracted proportional to the
number of its occurrences in the dictionary. Since typically LZ77
words already contains spaces, we do not include further spaces
separating them. It should be stressed as the structure of a
dictionary is affected by the size of LZ77 sliding window. In our case
we have typically adopted windows of 32768 characters, and, in a few
cases, of 65536 characters.

Below we present an excerpt of $400$ characters taken from an
artificial text (AT) having Melville's Moby Dick text as root.

\vspace{0.5cm} 

{\sffamily those boats round with at coneedallioundantic turneeling he
had Queequeg, man ."Tisheed the o corevolving se were by their fAhab
tcandle aed. Cthat the ive ing, head upon that can onge Sirare ce more
le in and for contrding to the nt him hat seemed ore, es; vacaknowt."
" it seemside delirirous from the gan . All ththe boats bedagain,
brightflesh, yourselfhe blacksmith's leg t. Mre?loft restoon }

\vspace{0.3cm}

As it is evident the meaning is completely lost and the only feature
of this text is to represent in a significant statistical way the
typical structures found in the original root text (i.e. the typical
subsequences of characters).

The case of sequences representing texts is interesting, and it is
worth spending a few words about it, since a clear definition of word
already exists in every language. In this case one could also define
\ti{natural} artificial texts (NAT). A NAT is obtained by
concatenating true words as extracted from a specific text written in
a certain language. Also in this case each word would be chosen
according to a probability proportional to the frequency of its
occurrence in the text. Just for comparison with the previous AT we
report an example of a natural artificial text built using real words
from the English dictionary taken randomly with a probability
proportional to their frequency of occurrence in Moby Dick's text.

\vspace{0.5cm} 

{\sffamily of Though sold, moody Bedford
opened white last on night; FRENCH unnecessary the charitable utterly
form submerged blood firm-seated barricade, and one likely keenly end,
sort was the to all what ship nine astern; Mr. and Rather by those of
downward dumb minute and are essential were baby the balancing right
there upon flag were months, equatorial whale's Greenland great
spouted know Delight, had } 

\vspace{0.5cm} 

We now describe how Artificial Texts can be effectively used for
recognition and classification purposes. First of all AT present
several positive features. They allow to define typical {\em words}
for generic sequences (not only for texts). Moreover for each original
text (or original sequence), one can construct an {\em ensemble} of
AT. This opens the way to the possibility of performing statistical
analysis by comparing the features of many AT all representative of
the same original root text. In this way it is possible to overcome
all the difficulties, discussed in the previous section, related to
the length of the strings analyzed. In fact it seems very plausible
that, once a certain ``reasonable'' AT size has been established, any
string can be well represented by a number of AT proportional to its
length. On the other hand one can construct AT by merging dictionaries
coming from different original texts: merging dictionaries extracted
from different texts all about the same subject or all written by the
same author. In this way the AT would play the role of an archetypal
text of that specific subject or that specific author~\cite{new-dict}.

\begin{figure}
\centerline{\psfig{figure=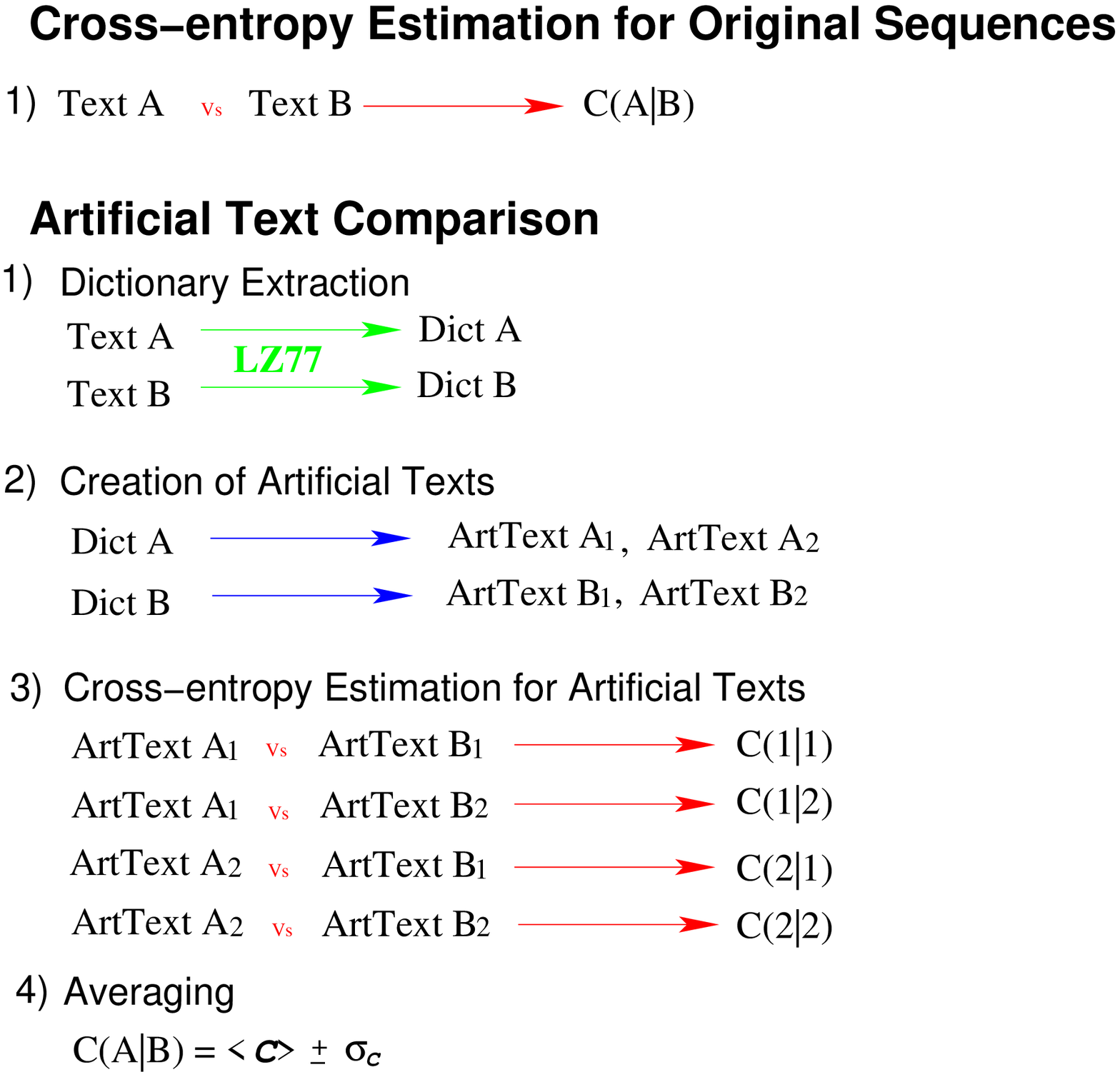,height=8cm,angle=0}}
\caption{{\bf Artificial Text Comparison (ATC) method:} This is the
scheme of the Artificial Text Comparison method. Instead of comparing
two original strings, several AT (two in figure) are created starting
from the dictionaries extracted from the original strings, and the
comparison is between pairs of AT. For each pair of AT coming from
different roots a cross-entropy value $C(i|j)$ is measured and the
cross-entropy between the root strings is obtained as the average
$<C>$ of all the $C(i|j)$. This method has the advantage of allowing
for an estimation of an error, $\sigma$, on the obtained value of the
cross-entropy $<C>$, as the standard deviation of the $C(i|j)$. From
the point of view of the ATC computational demand, point 1) simply
consists in the procedure of zipping the original files, that usually
requires few seconds, points 2) and 4) are of course negligible, while
point 3) is crucial. Obviously, in fact, the machine time requested
for the cross-entropy estimation grows as the square power of the
number of AT created (for fixed length of the AT).}
\label{f:atc}
\vspace{0.5cm}
\end{figure}

The possibility to construct many different AT all representative of
the same original sequence (or of a given source) allows for an
alternative way to estimate the self-entropy of a source (and
consequently the relative entropy between two sources as mentioned
above). The cross entropy between two AT corresponding to the same
source will give in fact directly an estimate of the self-entropy of
the source. This is an important point since in this way it is
possible to estimate the relative entropy and the distances between
two texts of the form proposed in eq.~\ref{distanza_nostra} in a
coherent framework. Finally, as it is shown in Figure~\ref{f:atc},
comparing many AT coming from the same two roots (or single root), we
can estimate a statistical error on the value of the cross-entropy
between the two roots.

With the help of AT we can then build a comparison scheme (Artificial
Text Comparison or ATC) (see figure~\ref{f:atc}) between sequences
whose validity will be checked in the following sections. This scheme
is very general since it can be applied to any kind of sequence
independently of the coding behind it. Moreover the generality of the
scheme comes from the fact that, by means of a re-definition of the
concept of word, we are able to extract subsequences from a generic
sequence using a deterministic algorithm (for instance LZ77) which
eliminates every arbitrariness (at least once the algorithm for the
dictionary extraction has been chosen). In the following sections we
shall discuss in detail how one can use AT for recognition and
classification purposes.

\section{Recognition of linguistic features}

Our first experiments are concerned with recognition of linguistic
features. Here we consider those situations in which we have a corpus
of \ti{known} texts and one unknown text $X$. We are interested here
in identifying the known text $A$ closest (according to some rule) to
the $X$ one. We then say that $X$, being similar to $A$, belongs to
the same group of $A$. This group can, for instance, be formed by all
the works of an author, and in that case we say that our method
attributed $X$ to that author. We now present results obtained in
experiments of language recognition and authorship attribution. After
having explained our experiments we will be able to make some more
comments on the criterion we adopted to set recognition and/or
attribution.

\subsection{Language recognition}

Suppose we are interested in the automatic recognition of the language
in which a given text $X$ is written. This case can be seen as a first
benchmark for our recognition technique. The procedure we use
considers a collection (a corpus), as large as possible, of texts in
different (known) languages: English, French, Italian, Tagalog
\ldots. We take an $X$ text to play the role of the unknown text whose
language has to be recognized, and the remaining $A_i$ texts of our
collection to form our background. We then measure the cross entropy
between our $X$ text and every $A_i$ with the procedure discussed in
section II. The text, among the $A_i$ group, with the smallest cross
entropy with the $X$ one, selects the language closest to the one of
the $X$ file, or exactly its language, if the collection of languages
contains this language. In our experiment we have considered in
particular a corpus of texts in $10$ official languages of the
European Union (UE)~\cite{UE}: Danish, Dutch, English, Finnish,
French, German, Italian, Portuguese, Spanish and Swedish. Using $10$
texts for each language we had a collection of $100$ texts. We have
obtained that for any single text the method has recognized the
language. This means that the text $A_i$ for which the cross entropy
with the unknown $X$ text was the smallest was a text written in the
same language. We found out also that if we ranked for each $X$ text
all the texts $A_i$ as a function of the cross entropy, all the texts
written in the same language of the unknown text were in the first
positions. This means that the recall, defined in the framework of
information retrieval as the ratio between the number of relevant
documents retrieved (independently of the position in the ranking) and
the total number of existing relevant documents, is maximal, i.e.
equal to one. The recognition of language works quite well for length
of the $X$ file as small as a few tens of characters.

\subsection{Authorship attribution}

Suppose now to be interested in the automatic recognition of the
author of a given text $X$. We shall consider, as before, a
collection, as large as possible, of texts of several (known) authors
all written in the same language of the unknown text and we shall look
for the text $A_i$ for which the cross entropy with the $X$ text is
minimum. In order to collect a certain statistics we have performed
the experiment using a corpus of $87$ different
texts~\cite{liberliber} of $11$ Italian authors, using for each run
one of the texts in the corpus as the unknown $X$ text. In a first
step we proceeded exactly as for language recognition, using the
actual texts. The results, shown in Table~\ref{ric-autore}, feature a
rate of success of roughly $93\%$. This rate is the ratio between the
number of texts whose author has been recognized (another text of the
same author was ranked as first) and the total number of texts
considered. There are of course fluctuations in the success rate for each
author and this has to be expected since the writing style is
something difficult to grasp and define; moreover it can vary a lot in
the production of a single author.

\begin{table}[bottom]
\begin{center}
\begin{tabular}{|c|c|c|c|c|}
\hline
{\centering {\sf {\bf AUTHOR}}} 
& {\centering {\sf Number of}} 
&{\centering {\sf Successes:}} 
&{\centering {\sf Successes:}} 
&{\centering {\sf Successes:}} \\
{\centering {\sf }} 
& {\centering {\sf texts}} 
&{\centering {\sf {Actual texts}}} 
&{\centering {\sf {ATC}}} 
&{\centering {\sf {NATC}}} \\
\hline
 {\centering {\sf Alighieri}}
&{\centering {\sf 5}}
&{\centering {\sf 5}}
&{\centering {\sf 5}}
&{\centering {\sf 5}}
\\\hline
 {\centering {\sf D'Annunzio}}
&{\centering {\sf 4}}
&{\centering {\sf 4}}
&{\centering {\sf 4}}
&{\centering {\sf 4}}
\\\hline
 {\centering {\sf Deledda}}
&{\centering {\sf 15}}
&{\centering {\sf 15}}
&{\centering {\sf 15}}
&{\centering {\sf 15}}
\\\hline
 {\centering {\sf Fogazzaro}}
&{\centering {\sf 5}} 
&{\centering {\sf 4}}
&{\centering {\sf 5}}
&{\centering {\sf 5}}
\\\hline
 {\centering {\sf Guicciardini}}
&{\centering {\sf 6}}
&{\centering {\sf 5}}
&{\centering {\sf 6}}
&{\centering {\sf 6}}
\\\hline
 {\centering {\sf Machiavelli}}
&{\centering {\sf 12}}
&{\centering {\sf 12}}
&{\centering {\sf 11}}
&{\centering {\sf 10}}
\\\hline
 {\centering {\sf Manzoni}}
&{\centering {\sf 4}}
&{\centering {\sf 3}}
&{\centering {\sf 4}}
&{\centering {\sf 4}}
\\\hline
 {\centering {\sf Pirandello}}
&{\centering {\sf 11}}
&{\centering {\sf 11}}
&{\centering {\sf 11}}
&{\centering {\sf 11}}
\\\hline
 {\centering {\sf Salgari}}
&{\centering {\sf 11}}
&{\centering {\sf 10}}
&{\centering {\sf 11}}
&{\centering {\sf 11}}
\\\hline
 {\centering {\sf Svevo}}
&{\centering {\sf 5}}
&{\centering {\sf 5}}
&{\centering {\sf 5}}
&{\centering {\sf 5}}
\\\hline
 {\centering {\sf Verga}}
&{\centering {\sf 9}}
&{\centering {\sf 7}}
&{\centering {\sf 9}}
&{\centering {\sf 9}}
\\\hline
 {\centering {\sf {\bf TOTALS}}}
&{\centering {\sf 87}}
&{\centering {\sf 81}}
&{\centering {\sf 86}}
&{\centering {\sf 85}}
\\\hline
\end{tabular}
\vspace{0.1cm}
\caption{{\bf Author recognition}: This table illustrates the results
for the experiments of author recognition. For each author we report
the number of different texts considered and a measure of success for
each of the three methods adopted. Labeled as successes are the
numbers of times another text of the same author was ranked in the
first position using the minimum cross-entropy criterion.}
\label{ric-autore}
\end{center}
\end{table}

We then proceeded analyzing the same corpus with the ATC method we
have discussed in the previous section. We extracted the dictionary
from each text, and built up our $87$ artificial texts (each one
$30000$ characters long). In each run of our experiment we chose one
artificial text to play the role of the text whose author was unknown
and the other $86$ to be our background. The result is
significant. We found that $86$ times on $87$ trials the author was
indeed recognized, i.e. the cross entropy between our unknown text and
at least another text of the right author was the smallest. This means
that the rate of success using artificial texts was of $98.8\%$. The
unrecognized text was {\em L'Asino} by Machiavelli, which was
attributed to Dante ({\em La Divina Commedia}), and, in fact, these
are both poetic texts; so it does not appear so strange thinking that
{\em L'Asino} is found to be in some way closer to the {\em Commedia}
rather than to {\em Il Principe}. A slightly different way to proceed
is the following. Instead of extracting an artificial text from each
actual text, we made a single artificial text, which we call the
\ti{author archetype}, for each author. To do this we simply joined
all the dictionaries of the author and then proceeded as before. In
this case we used actual works as unknown texts and author archetypes
as background. We obtained that $86$ out of $87$ unknown real texts
matched the right artificial author text, the one missing being again
{\em L'Asino}.

In order to investigate this mismatching further we exploited one of
the biggest advantages the ATC method can give if compared to the real
text comparison. While in real text comparison only one trial can be
made, ATC allows for creating an ensemble of different artificial
texts, and so more than one trial is possible. In our specific
case, however, $10$ ATC different trials performed both with artificial
texts and with author archetypes gave the same result, attributing
{\em L'Asino} to Dante. This can probably confirm our supposition that
the pattern of poetic register is very strong in this case. To be sure
that our $98.8\%$ rate of success was not due to a particular
fortuitous accident in our set of artificial texts, we repeated our
experiment with a corpus formed by $5$ artificial texts of each actual
text. This means that our collection was formed by $435$ texts. We
then proceeded in the usual way. Having our cross entropies between
the $5$ $X_n$ ($n=1,...,5$) artificial texts coming from the same root
$X$, and the remaining $430$ ATs, we first joined all the rankings
relative to these $X_n$. Thus we had $430 \times 5$ cross-entropies
between the AT extracted by the same root $X$ and the other AT of our
ensemble. We then averaged, for each root $A_i$, all the $25$ cross
entropies between an AT created from $X$ text and an AT extracted from
that $A_i$. In this way we obtained $86$ cross entropy values, and we
set authorship attribution using the usual minimum-criterion. We found
again that $86$ texts over $87$ were well attributed, {\em L'Asino}
being again mis-attributed.

This result shows that ATC is a robust method since it does not seem
to be strongly influenced by the particular set of artificial
texts. In particular, as we have discussed before, ATC allows for a
quantification of the error committed on the cross entropy
estimation. Defined as $\sigma_m$ the standard deviation estimated for
the $m^{th}$ cross-entropy, in a ranking in which the smallest cross
entropy value is the first one, we empirically observed these
relations:

\bea
\label{e:ling1} 
\frac{\sigma_1}{C_1} \simeq \frac{\sigma_2}{C_2} \simeq
\frac{\sigma_3}{C_3} \simeq 0.5 \%
\eea
\bea
(C_2-C_1) \simeq \sigma_1 \simeq \sigma_2.
\label{e:ling2}
\eea

The difference $C_2-C_1$ gives an indication of the level of
confidence of the results.  When this difference is of the order of
the standard deviation of $C_1$ and $C_2$, this is an indication that
the result for the attribution has an high level of confidence (at
least inside the corpus of reference files/texts considered).

Finally, in order to explore the possibility of using natural words,
we performed experiments with natural artificial texts. We call this
method Natural ATC or NATC. We built up 5 artificial texts for each
actual one using italian words instead of words extracted by
LZ77. Having these natural artificial texts we proceeded exactly as
before. We obtained that $85$ over $87$ texts where
recognized. Besides {\em L'Asino}, the other mismatch was the
\ti{Istorie Fiorentine} by Machiavelli that was set closest to
\ti{Storie Fiorentine dal 1378 al 1509} by Guicciardini. It seems
clear that the closeness of the subjects treated in the two texts
played a fundamental role in the attribution.

It is interesting trying some conjectures on why artificial texts made
up by LZ77 extracted dictionary worked better in our
experiment. Probably the main reason is that LZ77 very often puts some
correlation between characters and actual words by grouping them into
a single \ti{word}, while clearly this correlation does not exist
using natural words. In a text written to be read, words and/or
characters are correlated in a precise way, especially in some cases
(one of the most strict, but probably less significant, is ``.''
followed by a capital letter). These observations could maybe suggest
that LZ77 is able to capture correlations that are in some sense a
signature of an author, this signature being stronger (up to a certain
point, of course) than that of the subject of a particular text. On
the other hand this ability of keeping memory of correlations,
combined with the specificity of poetic register, could also explain
the apparent strength of poetic pattern that seems to emerge from our
experiments.

We have also performed some additional experiments on a corpus of
English texts. Results are shown in Table~\ref{ric-ingautore}. In this
corpus there were a few poetic texts which, as we could expect,
afflicted in some cases ATC. It is worth noting, in fact, that the
number of ATC failures is $7$, and in this case it's higher than that
of actual text comparisons, which is $4$. However, if we look
carefully we note that $4$ of this $7$ mismatches come from the $5$
Marlowe works present in our corpus. Among Marlowe's works only $1$ is
mis-attributed by actual text comparison, too. This peculiarity of
Marlowe roused our interest and we analyzed carefully Marlowe's
results. We found that one of the $4$ bad attributions was a poetic
text, \ti{Hero}, and was attributed to Spencer, while the remaining
$3$ unrecognized texts were all attributed to Shakespeare. Similar 
results were obtained using the NATC method which also does not allow
for a clear distinction between Marlowe and Shakespeare. Just as a
matter of curiosity, and without entering in the debate, we report
here that, among the many thesis on the real identity of Shakespeare,
there is one who claims Shakespeare was just a pseudonym used by
Marlowe to sign some of its works. The Marlowe Society embraces this
cause and has presented many works which should prove this theory, or
at least make it plausible (starting of course by confuting the
official date of death of Marlowe, 1593).

\begin{table}
\begin{center}
\begin{tabular}{|c|c|c|c|c|}
\hline
{\centering {\sf {\bf AUTHOR}}} 
& {\centering {\sf Number of}} 
&{\centering {\sf Successes:}} 
&{\centering {\sf Successes:}} 
&{\centering {\sf Successes:}} \\
{\centering {\sf }} 
& {\centering {\sf texts}} 
&{\centering {\sf Actual texts}} 
&{\centering {\sf ATC}} 
&{\centering {\sf NATC}} \\
\hline
 {\centering {\sf Bacon}}
&{\centering {\sf 6}}
&{\centering {\sf 6}}
&{\centering {\sf 6}}
&{\centering {\sf 6}}
\\\hline
 {\centering {\sf Brown}}
&{\centering {\sf 3}}
&{\centering {\sf 2}}
&{\centering {\sf 2}}
&{\centering {\sf 2}}
\\\hline
 {\centering {\sf Chaucer}}
&{\centering {\sf 6}}
&{\centering {\sf 6}}
&{\centering {\sf 6}}
&{\centering {\sf 6}}
\\\hline
 {\centering {\sf Marlowe}}
&{\centering {\sf 5}}
&{\centering {\sf 4}}
&{\centering {\sf 1}}
&{\centering {\sf 2}}
\\\hline
 {\centering {\sf Milton}}
&{\centering {\sf 8}}
&{\centering {\sf 8}}
&{\centering {\sf 7}}
&{\centering {\sf 7}}
\\\hline
 {\centering {\sf Shakespeare}}
&{\centering {\sf 37}}
&{\centering {\sf 37}}
&{\centering {\sf 37}}
&{\centering {\sf 37}}
\\\hline
 {\centering {\sf Spencer}}
&{\centering {\sf 7}}
&{\centering {\sf 5}}
&{\centering {\sf 6}}
&{\centering {\sf 5}}
\\\hline
 {\centering {\sf {\bf TOTALS}}}
&{\centering {\sf 72}}
&{\centering {\sf 68}}
&{\centering {\sf 65}}
&{\centering {\sf 65}}
\\\hline
\end{tabular}
\vspace{0.1cm}
\caption{{\bf Author recognition}: This table illustrates the
results for the experiments of author recognition. In this case ATC
results were afflicted by the presence in the corpus of a few poetic
texts that, as we have discussed, tend to recognize each others.}
\label{ric-ingautore}
\end{center}
\end{table}

Before concluding this section several remarks are in order concerning
our minimum cross-entropy method used to perform authorship
attribution. Our criterion has been that of saying that the $X$ should
be attributed to a given author if another work of this author is the
closest (in the cross-entropy ranking) to $X$. It can happen, and
sometimes this is the case, that the second-closest text to $X$
belongs to another author, different from the first. Said in other
words, in the ranking of relative entropies between the $X$ text and
all the other text of our corpus, works belonging to a given author
are far from clustering in the same zone of the ranking. This fact can
be easily explained with the large variety of features that can be
present in the production of an author. Dante, for instance, wrote
both poetry and prose, this latter both in Italian and Latin. In order
to take into account this non-homogeneity we decided to set authorship
by watching only at the closest text to the unknown one. In fact, for
what we have said, averaging or taking into account all the texts of
every author could introduce biases given to the heterogeneity in each
author's production. Our choice is then perfectly coherent with the
purpose of authorship attribution which is not to determine an
\ti{average} author of the unknown text, but who wrote that particular
text. The limit of this method is the assumption that if an author
wrote a text, then he is likely to have written a similar text, at
least with regard to structural or syntactic aspects. From our
experiments we can say, a posteriori, that this assumption does not
seem to be unrealistic.

A further remark concerns the fact that our results for authorship
attribution could only provide with some hints about the real
paternity of a text.  One cannot, in fact, never be sure that the
reference corpus contains at least one text of the unknown author. If
this is not the case we can only say that some works of a given author
resembles to the unknown text. On the other hand the method could be
highly effective when one has to decide among a limited and predefined
set of candidate authors: see for instance the {\em Wright-Wright}
problem~\cite{trebbi} and the {\em Grunberg-Van der Jagt} problem in
The Netherlands~\cite{dutch}.

From a general point of view, finally, it is important to remark that
the ATC method is of much greater interest than the NATC one. In fact,
even though in linguistic related problem the two methods give
comparable results, ATC can be used with every set of generic
sequences, while the NATC requires a precise definition of words in
the original strings.

\section{Self-consistent classification} 

In this section we are interested in the classification of large
corpora in situations where no a priori knowledge of corpora's
structure is given. Our method, mutuated by the phylogenetic analysis
of biological sequences~\cite{Cavalli-Edwards,Farris1,Fel1}, considers
the construction of a distance matrix, i.e. a matrix whose elements
are the distances between pairs of texts. Starting from the distance
matrix one can build a tree representation: phylogenetic
trees~\cite{Fel1}, spanning trees etc. With these trees a
classification is achieved by observing clusters that are supposed to
be formed by similar elements. The definition of a distance between
two sequences of characters has been discussed in section II.b.

\subsection{Author trees}

\begin{figure}
\centerline{\psfig{figure=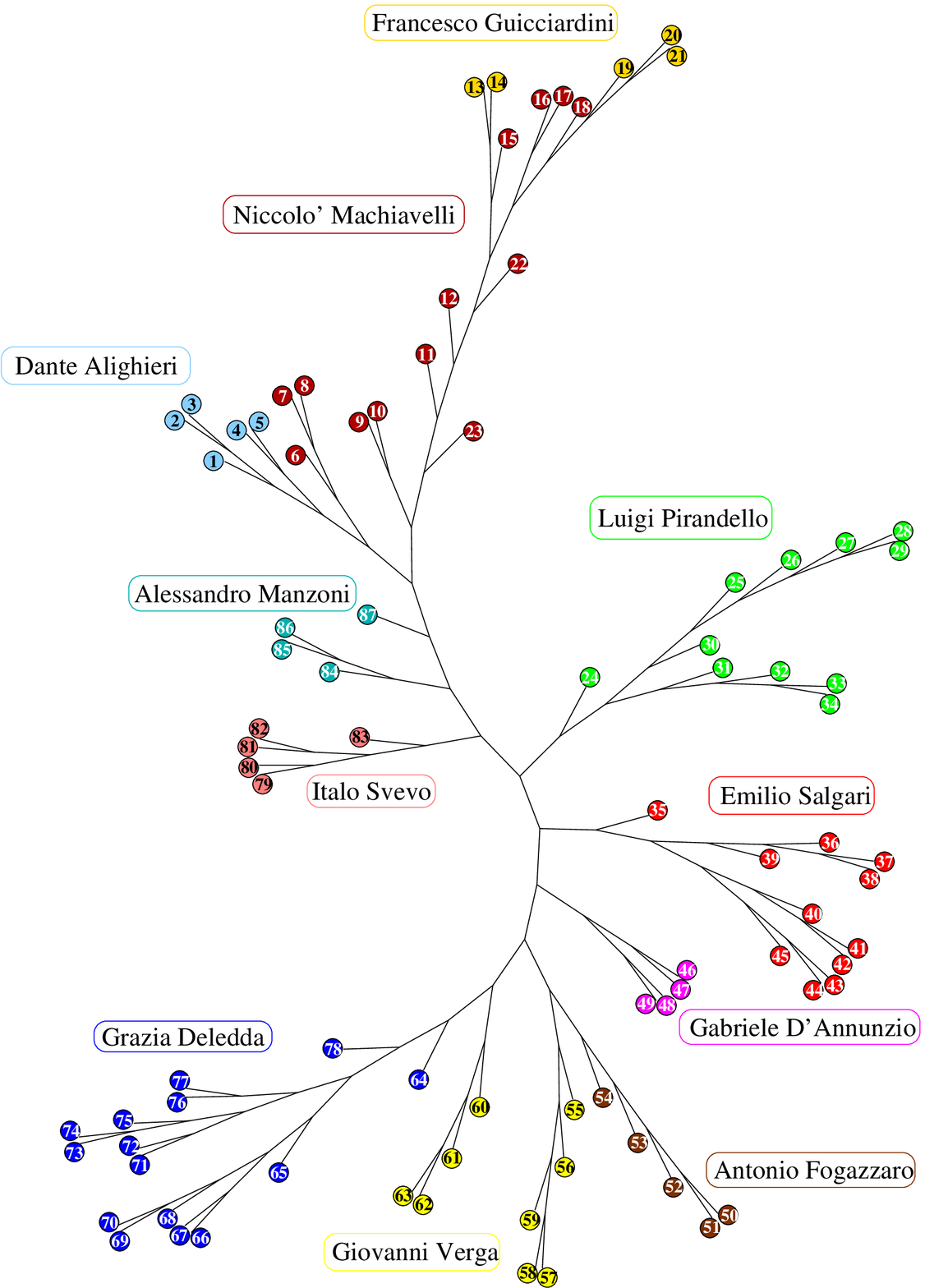,width=8cm,angle=0}}
\vspace{0.3cm}
\caption{ { \bf Italian Authors' tree:} Tree obtained with
Fitch-Margoliash algorithm using the $P$ pseudo-distance built from
ATC method for the corpus of Italian texts considered in sect.IV.b.
For sake of clarity in the representation we have chosen a constant
length for the distances between nodes and between nodes and leaves.}
\label{ita-albero}
\end{figure}

In our applications we used the Fitch-Margoliash
method~\cite{fitch-margo} of the package PhylIP (Phylogeny Inference
Package)~\cite{fel_Phylip} which basically constructs a tree by
minimizing the net disagreement between the matrix pairwise distances
and the distances measured on the tree. Similar results have been
obtained with the Neighbor algorithm~\cite{saitou-nei}. The first test
for our method consisted in analyzing with the Fitch-Margoliash
procedure the distance matrix obtained by the corpus of italian texts
used before for authorship attribution. Results are presented in
Figure~\ref{ita-albero}. As it can be seen works by the same author
tend to cluster quite well in the presented tree.

\subsection{Language trees}

The next step was applying our method in a less obvious context: that
of relationship between languages. Suppose to have a collection of
texts written in different languages. More precisely, imagine to have
a corpus containing several versions of the same text in different
languages, and to be interested in a classification of this corpus.
In order to have the largest possible corpus of texts in different
languages we have used: ``The Universal Declaration of Human
Rights''~\cite{unhchr} which sets the Guinness World Record for the
most translated document.

\begin{figure}
\centerline{\psfig{figure=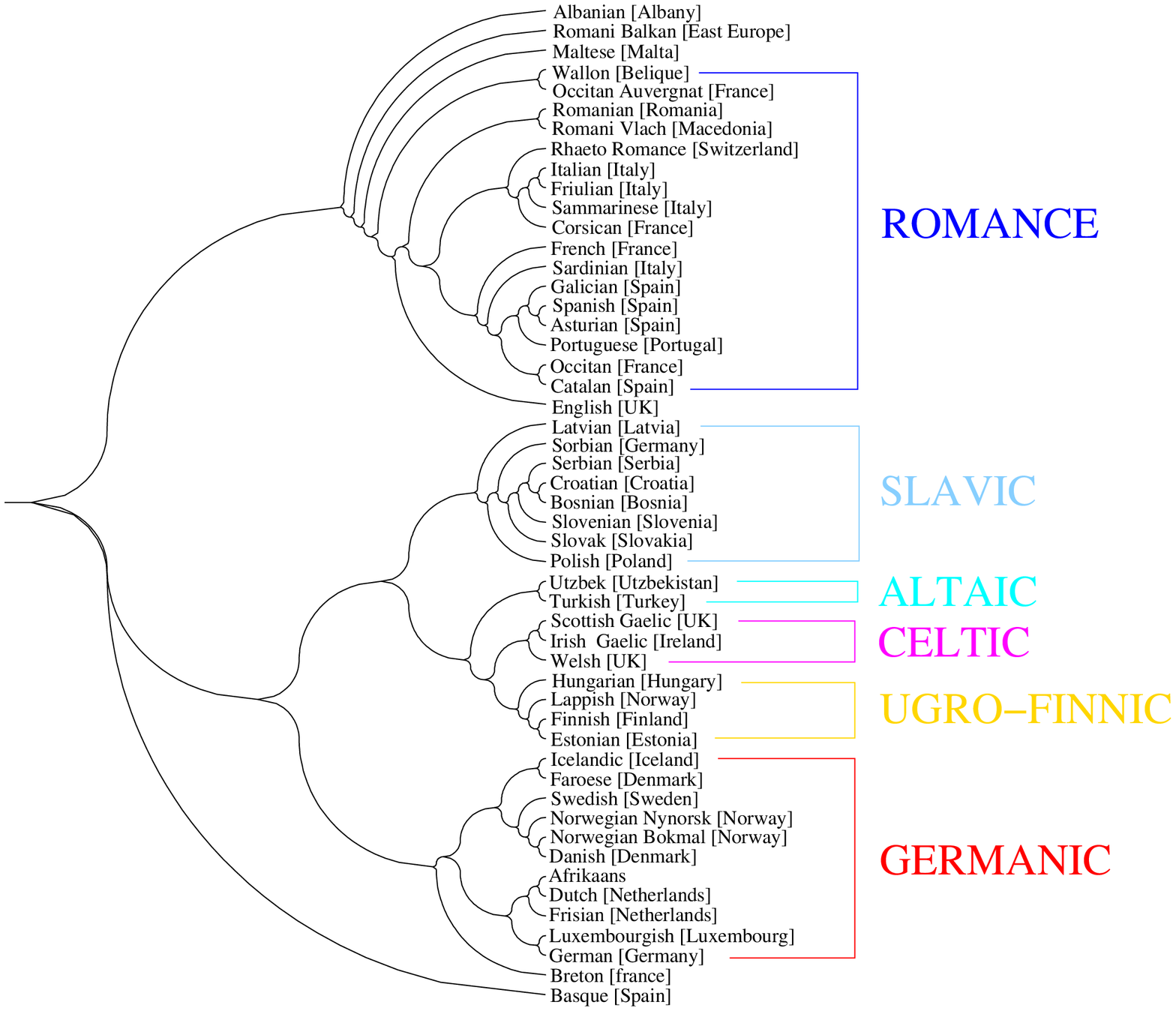,width=8cm,angle=90}}
\vspace{0.3cm}
\caption{{\bf Indo-european family language tree:} This figure
illustrates the phylogenetic-like tree constructed on the basis of
more than $50$ different versions of the ``The Universal Declaration
of Human Rights''. The tree is obtained using the Fitch-Margoliash
method applied to the symmetrical distance matrix based on the $R$
distance defined in sect. II.b built from ATC method. This tree
features essentially all the main linguistic groups of the
Euro-Asiatic continent (Romance, Celtic, Germanic, Ugro-Finnic,
Slavic, Baltic, Altaic), as well as few isolated languages as the
Maltese, typically considered an Afro-Asiatic language, and the
Basque, classified as a non-Indo-European language and whose origins
and relationships with other languages are uncertain.  The tree is
unrooted, i.e. it does not require any hypothesis about common
ancestors for the languages and it can not be used to infer
informations about common ancestors of the languages. For more details
see the text.The lengths of the paths between pairs of documents
measured along the tree branches are not proportional to the actual
distance between the documents.}
\label{lang-tree-eu}
\end{figure}

We proceeded here for our analysis exactly as for author trees. We
analyzed with the Fitch-Margoliash method~\cite{fitch-margo} the
distance matrix obtained using the Artificial Text Comparison method
with $5$ artificial texts for each real text. After averaging on the
Artificial Texts sharing the same root, we have built up the distance
matrix as discussed in section II.b. In Fig.~\ref{lang-tree-eu} we
show the tree obtained with the Fitch-Margoliash algorithm for over
$50$ languages widespread on the Euro-Asiatic continent. We can notice
that essentially all the main linguistic groups (Ethnologue
source~\cite{ethnologue}) are recognized: Romance, Celtic, Germanic,
Ugro-Finnic, Slavic, Baltic, Altaic. On the other hand one has
isolated languages as the Maltese, typically considered a Semitic
language because of its arabic base, and the Basque, a
non-Indo-European language whose origins and relationships with other
languages are uncertain. The results are also in good agreement with
those obtained by true sequences comparison reported in~\cite{bcl}
with a remarkable difference concerning the Ugro-Finnic group here
fully recognized, while with true texts Hungarian was put a little
apart.

After the publication of our tree in ~\cite{bcl} a similar tree, using
the same dataset, has been proposed in~\cite{li_similarity} using
$NCD(x,y)$ (see Sect. IIB) estimated with gzip.

It is important to stress how these trees are not intended to
reproduce the current trends in the reconstruction of genetic
relations among languages. They are clearly biased by the fact of
using entire modern texts for their construction. In the
reconstruction of genetic relationships among languages one is
typically faced with the problem of distinguishing {\em vertical}
(i.e. the passage of information from parent languages to child
languages) from {\em horizontal} transmission (i.e. which includes all
the other pathways in which two languages interact). This is the main
problem of lexicostatistics and glottochronology~\cite{renfrew} and
the most widely used method is that of the so-called Swadesh
$100$-words lists~\cite{swadesh}. The main idea is that of comparing
languages by comparing lists of so-called basic words. These lists
only include the so-called cognate words ignoring as much as possible
horizontal borrowings of words between languages. It is clear now how
an obvious source of bias in our results is represented by the fact of
non-having performed any selection of words to be compared. It turns
out then that in our trees English is closer to Romance languages
simply because almost $50\%$ of English vocabulary has been borrowed
from French. These borrowings should be expunged if one is interested
in reconstructing the actual genetic relationships between
languages. Work is presently in progress in order to merge Swadesh
list techniques with our methods~\cite{starostin}.

\section{Discussion and Conclusions} 

We have presented here a class of methods, based on the LZ77
compression algorithm, for information extraction and automatic
categorization of generic sequences of characters. The essential
ingredient of these methods is the definition and the measure of a
remoteness and of a distance between pairs of sequences of
characters. In this context we have introduced in particular the
notion of {\em dictionary} of a sequence and of {\em Artificial Text}
(or {\em Artificial Sequence}) and we have implemented these new tools
in an information extraction scheme (ATC) that allows to overcome
several difficulties arising in the comparison of sequences.

With these tools in our hands, we have focused our attention on
several applications to textual corpora in several languages, since in
this context it is particularly easy to judge experimental results. 
We have at first shown that dictionaries are intrinsically interesting
and that they contain relevant signatures of the texts they are
extracted from. 
Then in a first series of experiments we have shown how we can determine, and
then extract, some {\em semantic} attributes of an unknown text (its
language, author or subject). We have also shown that comparing
artificial texts, instead of
actual sequences, gives better results in most of these situations. In
the linguistic context, moreover, we have been able to define natural
artificial texts (NAT) exploiting the presence of natural language words in the
analyzed texts. Results from experiments indicate that this
additional information does not produce any advantage, i.e. the
NAT comparison (NATC) and ATC yield to the same results. 
However, the question is not whether NATC performs
better than ATC. From a general point of view, in fact, the ATC
method is of much greater interest with respect to 
the NATC one. In fact, while in linguistic related problems the
two methods equally perform, in many cases NATC are impossible to
construct because outside linguistics there is no precise definition
of word. On the other hand the fact that ATC and NATC perform at least
equally well in linguistics motivated problems, is a good news because
one can reasonably infer that the situation will not change drastically
in situations where NATC will not be available anymore.

A slightly different application of our method is that of the
self-consistent classification of a corpus of sequences. In this case
we do not need any information about the corpus, but we are interested
in observing the self organization that arises from the knowledge of a
matrix of distances between pairs of elements. A good way to represent
this structure can be obtained using phylogenetic algorithms to build
a tree representation of the considered corpus. In this paper we have
shown how the self-organized structures observed in these trees are
related to the semantic attributes of the considered texts.

Finally, it is worth stressing once again the high versatility and
generality of our method that applies to any kind of corpora of
character strings independently of the type of coding behind them:
texts, symbolic dynamics of dynamical systems, time series, genetic
sequences, etc. These features could be potentially very important for
fields where the human intuition can fail: genomics, geological time
series, stock market data, medical monitoring, etc.

{\large Acknowledgments}: The authors are indebted with Dario
Benedetto with whom part of this work has been completed. The authors
are grateful to Valentina Alfi, Luigi Luca Cavalli-Sforza, Mirko Degli
Esposti, David Gomez, Giorgio Parisi, Luciano Pietronero, Andrea
Puglisi, Angelo Vulpiani, William S. Wang for very enlightening
discussions.

\end{document}